# Iris: A Conversational Agent for Complex Tasks

**Ethan Fast, Binbin Chen, Julia Mendelsohn, Jonathan Bassen, Michael Bernstein**
Stanford University
{ethaen, bchen45, jmendels, jbassen}@stanford.edu, msb@cs.stanford.edu

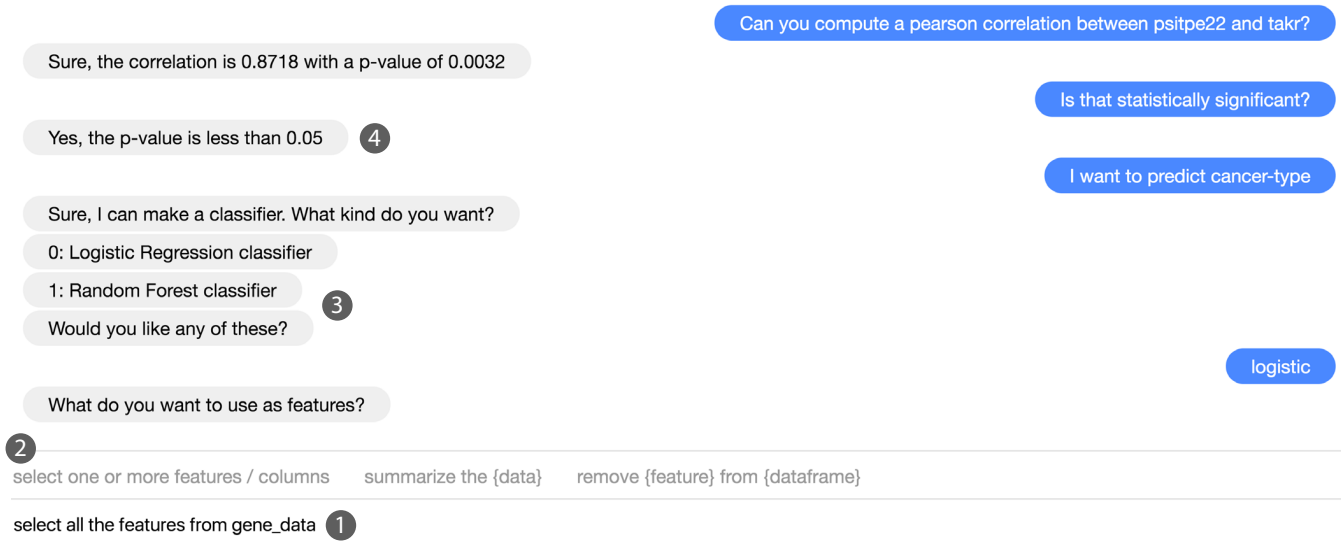

Figure 1: Iris is a conversational agent that helps users with data science tasks. (1) Users interact with Iris through natural language requests and (2) the system responds with real-time feedback on the command the request will trigger. Once a command is triggered, Iris (3) converses with users to resolve arguments, which may also be the result of a (1) new command or (4) previous conversation.

**ABSTRACT**
Today's conversational agents are restricted to simple stand-alone commands. In this paper, we present Iris, an agent that draws on human conversational strategies to *combine commands*, allowing it to perform more complex tasks that it has not been explicitly designed to support: for example, composing one command to "plot a histogram" with another to first "log-transform the data". To enable this complexity, we introduce a domain specific language that transforms commands into automata that Iris can compose, sequence, and execute dynamically by interacting with a user through natural language, as well as a conversational type system that manages what kinds of commands can be combined. We have designed Iris to help users with data science tasks, a domain that requires support for command combination. In evaluation, we find that data scientists complete a predictive modeling task significantly faster (2.6 times speedup) with Iris than a modern non-conversational programming environment. Iris supports the same kinds of commands as today's agents, but empowers users to weave together these commands to accomplish complex goals.

**ACM Classification Keywords**
H.5.2 Information Interfaces and Presentation: Natural Language

**INTRODUCTION**
For decades, the promise of computers that communicate with us through natural language has been depicted in works of science fiction and driven research agendas in artificial intelligence (AI) and human-computer interaction (HCI). As early as 1964, Joseph Weizenbaum demonstrated how a computer program could hold open-ended conversations using a large set of pattern matching rules [45]. Terry Winograd later developed a more sophisticated program that could act upon natural language requests within a simplified "blocks world" [46]. In recent years, virtual assistants such as Siri and Cortana have increasingly applied conversational agents to real-world problems, such as finding local restaurants and scheduling calendar appointments [24].

Speak to these conversational agents as you would to a colleague or graduate student, however, and it becomes clear that they have serious limitations. When you ask a conscientious graduate student to "check for significant differences between conditions", they might reply, "what kind of differences?" And you would say, "statistical differences, you know, through a t-test." And they might ask, "you're talking about income for the high- and low-education conditions?" And you would say, "that's right, although they are skewed, so you should log-transform them first." The end result of this conversation is that the graduate student would run a t-test to check for differences between the log-transforms of income data for high- and low-education testing populations. Try something like this with Siri, on the other hand (assuming Siri understood t-tests and log-transforms), and it would respond with "I don't know what you mean".

Why does Siri fail? The root of the issue is that Siri (and similar agents) employ a simple model of conversation. While humans engage in complex conversations, interactions with today's agents can be described by the following algorithm: (1) determine which command a user wants, (2) ask for any missing arguments, and (3) execute it. In other words, every conversation you have with an agent today is equivalent to *executing a standalone command*. These standalone commands are effective for simple goals such as setting a timer or finding directions, but they restrict agents to tasks that a system has been explicitly designed to handle.

Linguistic theory suggests how conversational agents can overcome this limitation. People do not restrict themselves to standalone commands: we build meaning through *combinations* of them [16, 28]. We combine commands, for example, by nesting conversations. When a colleague says, "should I do a t-test or Mann-Whitney U?", we might respond with a new conversation, "well, is the data normally distributed?". Similarly, we combine commands through references to previous ones: after asking "did you take the log-transform?", we might say, "what is the variance of *that*?"

In this paper, we explore an architecture for conversational agents that enables complex tasks through command combination. This architecture allows users to interact with an agent to *compose* commands (for example, to make a histogram of log-transformed data) or apply commands in *sequence* (for example, to extract a list of column names and then run a t-test on each). We show how this architecture allows users to execute complex commands without predefined support in the system by weaving together multiple lower-level commands.

We introduce two technical contributions that enable conversational agents to combine commands. When agents only deal with one command at a time, the logic managing each command can execute in isolation. When composing commands, however, this logic must expand and modify itself at runtime to combine with other commands. Our first contribution is a domain specific language (DSL) that transforms commands into an automata-based programming model, allowing our system to support dynamic transformations.

When allowing agents to combine commands, agents must understand which commands *can be* combined. For example, it is reasonable to ask an agent to "make a histogram" of log-transformed data but not of a t-test. An agent that allows users to combine commands should provide guardrails that prevent failure when users ask it to do something that would result in an error or when it misinterprets a user request. Our second contribution is a *conversational type system* that allows agents to infer which commands can be combined and gracefully handle type mismatches.

We showcase this architecture in *Iris*, a conversational agent that helps users with data science tasks (Figure 1). Data science represents a domain with complex tasks that standalone commands cannot support [18]. To interact with Iris, you type natural language requests into a chat window (1.1) and receive real-time feedback on what command your request will trigger (1.2). When you enter a command, Iris converses with you to resolve its arguments (1.3), which you may populate by calling new commands, a form of conversational *composition* (1.1). You can also use the results of previous commands by storing them in named variables or referencing previous command results, forms of *sequencing* (1.4). If you are an expert user, Iris exposes an API that allows you to extend it with new commands (Figure 6).

Human language consists of far more than single-command utterances. Iris does not attempt to cover all of human language, and much linguistic complexity is beyond today's natural language processing (NLP). Instead, by creating agents that combine commands, we envision support for a class of interactions with increased expressivity and complexity.

## CONVERSATION ANALYSIS THEORY

How do people speak to one another? We draw on human conversational strategies in order to inspire Iris's design. In particular, we leverage conversation analysis (CA), a data-driven theory that is influential in sociolinguistics and discursive psychology [16, 12]. CA helps us explain what is missing in today's conversational agents and identify the human behaviors we would like these agents to emulate.

CA theory models all conversations through a basic unit called an *adjacency pair*: a pair of statements spoken by two conversational participants [16]. These pairs are typed with labels such as greeting-greeting or question-answer. Iris's *conversational type system* similarly constrains the responses that users can provide to execute commands.

More complex conversations can be described by joining adjacency pairs through expansions: (1) *pre-expansions* prepend an adjacency pair that allows one speaker to perform a preliminary request, for example, a question about availability that precedes an invitation; (2) *insert expansions* nest an adjacency pair within a conversation that allows one speaker to resolve other issues before continuing, for example, asking what time a World Cup game airs before answering a question about when to schedule a meeting; (3) *post-expansions* append an adjacency pair that allows one speaker to signal acknowledgment, for example, "got it".

Insert expansions are particularly relevant to conversational agents because they can redirect the substance of a conversation. Today's agents often leverage a form of insert expansion that linguists call the *clarification request* [28]. Humans use these requests to clarify the meaning of previous statements, which makes them the most commonly observed form of insert expansion. Agents (including Iris) use clarification requests in a similar but more constrained way, such as confirming arguments before execution: for example, "By Elena, did you mean Elena Ferrante?" (Table 1).

Unlike today's conversational agents, Iris supports a powerful form of insert expansion called the *dependent question*: a question that depends on the answer to some subsequent request. Humans use dependent questions when the answer to one question ("Will you be going to the game tonight?") depends on the answer to another ("Who else will be there?"). Dependent questions are the second most common insert expansion observed in human conversation [28]. Iris supports these questions by composing commands. For example, when Iris asks, "What array would you like to use?", a user can answer with, "Generate a new array from the normal distribution", and the resulting value will resolve the original question. We model these insert expansions through command *composition* (Table 1).

| | Example | Theory | Siri, etc. | Iris |
|---|---|---|---|---|
| Execute Command | Set an alarm for 12pm. Okay, it is set. | Speech acts | ✓ | ✓ |
| Resolve Arguments | Set an alarm. When do you want to set an alarm? | Clarification Request | ✓ | ✓ |
| Extract Arguments | Make a reservation at *Oren's* for *5pm*. Okay, making the reservation. | NA | ✓ | ✓ |
| Compose Commands | Schedule a meeting with Elena. Sure, for when? The day I get back from Australia. Okay, setting the meeting. | Dependent Question | Hard-coded | ✓ |
| Sequence Commands | Who is my wife? Elena Ferrante. Call her. Okay, calling Elena. | Anaphora | Hard-coded | ✓ |

Table 1: Interactions that today's conversational agents have been designed to support. Iris enables broader support for *composition*, calling one command within another command, and *sequencing*, using the results of previous commands in a new command.

Finally, the linguistic idea of *anaphora* describes expressions of language that depend on previous expressions [37]. For Iris, anaphoric expressions are values produced by some previous command that are necessary to the execution of the present command—in other words, they enable command sequencing (Table 1). Iris supports such expressions through named variables and a model of pronoun co-reference.

In sum, Iris draws on the concepts of insert expansions, dependent questions, and anaphora to support human conversational behaviors that are largely missing in today's systems.

## TODAY'S CONVERSATIONAL AGENTS

Iris is inspired by commercial agents such as Siri and their open source variants [15]. We examine the interactions these agents are designed to support in Table 1. Iris has much in common with existing conversational agents. All systems have the ability to *execute* standalone commands, such as "set a timer" or "generate a random number"; *extract arguments* from a natural language user query, for example, parsing the contact Oyeyemi from "send Oyeyemi an email"; and *resolve arguments* by asking follow-up questions, for example, responding to the request "schedule a meeting" with "when would you like me to set it?"

One important advance of Iris over these agents is support for *composition* (i.e., dependent questions). For example, when Iris asks you a question, such as "what classifier should I use?" you can respond with a request that initiates a new conversation such as "make a new logistic regression model". The result of that nested conversation will then pass back to the initial request. Today's systems support these interactions only in special cases. For example, when Siri asks you, "What's the date and time of your event?", you can respond with the time of another event such as "when do I get back from Australia?", but not with other commands that would work independently, such as "when is Elena's birthday?".

Similarly, Iris supports the *sequencing* of any of its commands (i.e., anaphora resolution). For example, you might ask Iris to "get the petal-length column from data.csv", then to "take the mean of that column". Existing agents support these interactions only when they have been hard-coded. For example, after you have successfully looked up a restaurant with Siri, you can ask "what time does it open?". But after Siri has returned a list of tweets via "search Twitter for tweets about Donald Trump", you cannot ask "show me the first one" even though that command works with other results, such as when navigating a list of restaurants.

In other words, *today's conversational agents only support composition and sequencing when these behaviors have been hard-coded into the system for specific commands*, so most commands in these systems cannot be stitched together. By fully supporting a user's ability of combine commands—through composition and sequencing—we can unlock a much more complex set of tasks.

## RELATED WORK

Iris belongs to a line of HCI systems that map natural language to underlying functionality such as commands, code snippets, and APIs. Query-feature graphs set the foundation for these methods, connecting user requests with system commands in an image editing application [10]. Others have since extended this approach: for example, using word embedding models to connect user vocabulary with system keywords [1, 5] or programming syntax [38], and statistical language models to predict functions a user wants to call [9, 26, 6]. All of these systems solve the vocabulary problem [10] through statistical models that map natural language to the domain language of a system. While Iris takes a similar tack, it extends user requests through conversation, allowing users to combine atoms of functionality—through composition and sequencing–into complex commands.

When designing for natural language interfaces, systems must manage the ambiguity of user language. DataTone and PixelTone provide guidance here, illustrating how systems can be designed to surface decisions about ambiguity [11] (at a point where these decisions matter), and constrain some of these decisions by direct manipulation [21] or other modalities [42]. Iris combines these ideas with an understanding of how humans resolve ambiguity (e.g., through clarification requests [28]) to manage the execution of user commands.

Systems that interact with users through speech have unique psychological constraints [36], such as the tendency of users to anthropomorphize them [32]. To better design for these constraints, other work in HCI has sought to model how humans communicate through theories such as the language/action perspective [47] and apply these theories to agents [29, 43]. Such theories have centered around individual speech acts, which describe how language relates to the world [39]. Iris expands on this perspective by capturing the interleaving of multiple speech acts (e.g. through insert expansions) as described by CA theory [16]. A second class of work has turned a practical lens on the challenges of implementing speech interfaces, providing tools for rapid prototyping [20], triggering user queries [48], or extending the limits of machine reasoning with a crowd [4, 22, 23]. One of Iris's goals is similarly to provide a framework (through its DSL and command API) that allows others to more quickly design conversational agents.

Iris also draws on insights from tools designed to help with data science tasks [34, 44, 41]. Wrangler, for example, combines spreadsheet visualizations with natural language command descriptions to help users manipulate data [17]. Iris leverages similar natural language descriptions in the hints it displays as a user formulates a command. Other tools such as Burrito and Variolite are oriented more towards organizing data science code for reuse [13, 19]. Iris also aims to enable reusable code, but contributes a different perspective: by wrapping high-level functions in natural language, users can redeploy these functions in future conversations.

Finally, Iris is inspired by a history of work in dialogue systems. In the AI literature, these systems fall along two primary axes: modular versus end-to-end, and structured versus unstructured. Modular systems, like Siri, train multiple models to support a set of tasks [15], while end-to-end systems train a single learning algorithm on dialogue data [40]. Similarly, structured systems assume a logical representation for the information exchanged in conversation (for example, slot filling systems [31]), which unstructured systems do not require [27]. Iris is a modular, structured system in that it is trained for a concrete set of tasks and contains structural information provided by its conversational type system. To our knowledge, Iris is the first dialogue system to enable the general combination of such commands through conversation.

## SCENARIO

To motivate how combining commands through sequencing and composition can empower interaction with a conversational agent, we present a scenario in which we use Iris to replicate work analyzing linguistic signals of dogmatism (intractability of opinion) in social media [8]. For this scenario, Iris is seeded with a large set of statistical commands based on the *scipy.stats* and *sklearn* libraries. We have also implemented commands that enable text analysis through popular lexicons [35, 7].

The original study collected dogmatism labels for Reddit posts through a crowdsourcing task. The dataset has two columns: *post*, the text of a post; and *score*, a dogmatism label between 0 (non-dogmatic) and 15 (dogmatic). We begin the scenario with the data loaded into the `dogmatism_data`, a collection indexed on these column names.

## Linguistic Analysis of Dogmatism

To discover linguistic features associated with dogmatic posts, we first need to divide the posts into two populations: dogmatic and non-dogmatic. The original study considered posts in the highest quartile (top 25%) as dogmatic and those in the lowest quartile (bottom 25%) as non-dogmatic.

### Composing Commands

We type "quartiles" into Iris and the hint box fills with `compute quartiles for an {array}`. This looks like what we want, so we trigger that command. We did not specify any arguments, so Iris asks: "What is the array you want to analyze?"

We want the score column in the original CSV file, stored in the collection `dogmatism_data`. The most direct way to accomplish this is to ask Iris to use "the score column in dogmatism_data", which would trigger a command to extract that column from the collection. This interaction is possible because Iris supports composition through insert expansions, calling one command (data selection) to fill in the argument for another (compute quartiles). Iris responds: "Q1 is from 2.0 to 7.0, Q2 is from 7.0 to 9.0, Q3 is from 9.0 to 12.0, and Q4 is from 12.0 to 15.0".

### Type System Guardrails

Instead of telling Iris to extract the score column, what if we had become confused and asked to use "dogmatism_data" to compute the quartiles? Because `dogmatism_data` is not an Array type, Iris would respond, "I need an Array but you've given me a Collection. Would you like to use a column from the Collection as an array?", presenting a list of columns we might select from. In this scenario, Iris's *conversational type system* allows it to convert the type the user provided (a Collection) into the type the command needs (an Array) though a set of conversion functions associated with each type, similar to duck typing in other languages (Figure 2B). These conversions allow Iris to fill in gaps between a user's understanding of a request and the system's underlying representation of data.

### Saving and Sequencing Commands

Next we need to select the highest and lowest quartiles of posts: those with scores less than 7 (non-dogmatic) and those with scores greater than 12 (dogmatic). We type "give me rows in dogmatism_data with score less than 7" and verify this will trigger the command `filter collection {dogmatism_data} with {score} column less than {7}`. Iris was able to automatically extract the arguments from our request, so we see them captured in brackets by the hint. We enter the command and Iris responds by displaying the filtered collection in the chat window (Figure 2C).

We want to reuse the value Iris just calculated, so we ask to "save that as dogmatic posts", which triggers `save {that} as {dogmatic_posts}`. A new variable named `dogmatic_posts` appears in the right side-bar. This interaction is possible because Iris can sequence commands, referencing the result of the previous command (via "that") in the current conversa-tion. Through a similar interaction, we save posts with scores greater than 12 in `non_dogmatic_posts`.

Now we would like to test these collections of posts for differences in linguistic features. To replicate the original paper, we will use LIWC (Linguistic Inquiry and Word Count) [35], a popular tool for computational social science embedded into Iris. LIWC analyzes text for signals across many linguistic categories, such as dominance, anger, or positive emotion. We tell Iris to "run an analysis using LIWC", which triggers the command `liwc analysis on {documents}`. Iris asks, "Where are the documents?"

We select the column of documents by asking for "the post column from dogmatic posts". Iris runs the analysis and returns a collection indexed on word counts for each of LIWC's linguistic categories. We save these category scores in `dogmatic_liwc`, and then repeat this process for the content of non-dogmatic posts and save them in `non_dogmatic_liwc`.

*Learning from User Language*
We can now test for linguistic signals that are different between dogmatic and non-dogmatic posts. We type "run Mann-Whitney tests between the columns in dogmatic_liwc and non_dogmatic liwc", which triggers `compute {statistical test} between {dogmatic_liwc} and {non_dogmatic_liwc}`.

In this case, Iris does not immediately understand which statistical test we have asked for, so it responds, "Sure, I can run statistical tests between two data collections. What test would you like to run?", along with a set of options. We select Mann-Whitney U, and Iris connects this with our use of "Mann-Whitney" in the original request, learning a new way to extract this option from input text that it will *remember in the future*. Iris then executes the command to generate a collection of test statistics, which we save in `dogmatism_stats`.

*Introspecting Iris Commands*
The original study corrected these test statistics via the Holmes method. Has Iris already done so? We ask, "can you tell me more about what you did?", and Iris responds with help text about the most recently executed command, indicating the statistics are uncorrected. This help text is also accessible at any time within a conversation. Alternatively, we might ask Iris to see a command's underlying Python code in scikit. Since the statistics have not been corrected, we ask Iris to "apply Holmes correction to dogmatism stats" and it applies the correction. We then "select significant statistics from those data" and "save that in final stats".

*Displaying Plots and Non-textual Data*
So, what are the statistically significant differences between dogmatic and non-dogmatic posts? We ask Iris to "plot odds ratios for the final_stats" to generate Figure 2D. In line with the original study, we see dogmatic associations for *swearing*, *negative sentiment*, and *sexual language*, and non-dogmatic associations for *first-person pronouns* and *past tense*.

*Saving Conversations as Code*
Finally, we want to save this analysis as Python code for replication. We ask Iris to "export this conversation as a script". This translates an abstract syntax tree (AST) representation of the current conversation into a Python program (Figure 5).

**A**: Composing commands

> find quartiles

Sure, I can compute quartiles
What array do you want to use?

> the score column in dogmatism_data

Sure, I'm using this array:
`[ 15.  15.  15. ...,   3.   3.   2.]`
Q1 is 2.0 to 7.0, Q2 is 7.0 to 9.0, Q3 is 9.0 to 12.0, and Q4 is 12.0 to 15.0

**B**: Conversational type system

> find quartiles

Sure, I can compute quartiles
What array do you want to use?

> dogmatism_data

I need an Array but you've given me a Collection. Would you like to use a column from the Collection as an array?

> yes

Here are the columns in that collection:
`['post' 'score' 'category']`
Which column would you like to select to use as an array?

> score

Great, I'm using score
Q1 is 2.0 to 7.0, Q2 is 7.0 to 9.0, Q3 is 9.0 to 12.0, and Q4 is 12.0 to 15.0

**C**: Sequencing commands

> filter dogmatism_data with score > 12

Here is the new collection:
`<Collection: [post, score, category]>`

> save that as dogmatic_posts

Saving as 'dogmatic_posts'
`<Collection: [post, score, category]>`

> can you show me the post column in dogmatic_posts?

Sure, here is the 'post' column:
`['>$150B every year...campaign?,`
` ...`
` 'cats cost money...lol cats']`

**D**: Displaying non-textual data

> plot odds ratios for final_stats

Sure, I can plot the odds ratios:

[bar chart: dogmatic_liwc — swear, sexual, anger, relig, death; odds associated with dogmatic_liwc, 0 to 6]

[bar chart: non_dogmatic_liwc — i, home, feel, family, focuspast; odds associated with non_dogmatic_liwc, 0.0 to 2.5]

**Figure 2: Iris replicates work analyzing dogmatism in social media: (A) the user composes a data selection command with a command to find quartiles; (B) the users saves a result, then uses it later; (C) Iris uses its conversational type system to provide guidance when the user misinterprets a request; (D) the user plots significant LIWC categories**

In this scenario, we showed how composition and sequencing enable users to accomplish complex tasks, such as computing statistical relationships. We also demonstrated how Iris's conversational type system provides guardrails to user interactions and can perform conversions automatically to allow a more seamless combination of commands.

## IRIS

Iris is a conversational agent that helps users with data science tasks. In this section, we describe how we built Iris, with emphasis on its compositional architecture, conversational type system, and API for command creation.

We present a high level overview of the Iris architecture in Figure 3. The entry point into interactions with Iris is a statistical model that maps user input onto commands. These commands require arguments that Iris either extracts from a request or converses with a user to resolve. Unlike today's agents, users can answer an argument request by executing another command (composition) or referencing the output of a previous conversation (sequencing) to form compound commands of arbitrary depth.

### A Programming Model for Conversation

How do we represent conversations as functions? To run a command, an agent needs to keep track of which command the user wants to execute, what arguments the user has provided so far in the conversation, and what arguments it still needs to request from the user. One representation adopted for this task by previous systems is a state machine [47], which loops over a set of questions associated with a command until it has enough information to execute it. To enable command composition, however, state machines must be able to *dynamically modify their transitions* to pass control of a conversation to new commands at runtime, and support a memory model that includes *scope* to bind arguments in nested conversations. To meet these needs, we have created a DSL that retains the simplicity of authoring commands in isolation but can support dynamic modification through transformation into an automata-based programming model.

In automata-based programming, execution of a program is broken down into explicit states of computation [14]. These states encapsulate a section of code to be run, after which the program will transition (much like a finite state machine) to the next state of computation. The DSL we have designed extends this idea with states that explicitly model interaction through conversation. At every step of computation, a state may output or receive input from a user, and any input it receives is passed as an argument to the section of code to be run for that computation. Each step of computation then returns a reference to the next state to be run. Together, these properties allows Iris to encode complex interactions through a set of linked, modular states that govern turn-taking with a user. In terms of conversation analysis, each state defines an adjacency pair.

More concretely, our DSL contains APIs for defining conversational automata, generating scopes over specified transitions (e.g., composed commands) to prevent binding conflicts between variables in memory, and setting transitions between automata dynamically. This DSL is open source as part of Iris, and it forms the basis for all the system's conversational abstractions.

### The Structure of Iris Commands

Iris commands are building blocks that users can combine to accomplish complex tasks. These commands are defined by a wrapper over Python functions that extends them with conversational affordances such as: (1) what user requests trigger the function, (2) what types of arguments the function requires, and how Iris should interact with a user to resolve them, (3) how Iris can extract arguments from example requests, and (4) how Iris should communicate the return value to the user. In sum, Iris commands are functions annotated with conversational metadata.

We present an example Iris command in Figure 4. The *title* field describes how Iris will reference the command in conversation. The *examples* provide initial grounding for what user requests will trigger the command and how to extract arguments. The *argument_types* connect arguments with types that dictate how Iris should converse with a user to resolve them. The *explanation* function defines how the return value of a command should be presented to the user.

Iris commands inherit from a base class (`IrisCommand`) that transforms them into automata via the DSL (Figure 3). An outermost automaton *extracts arguments* from a user's request and binds them to a command.

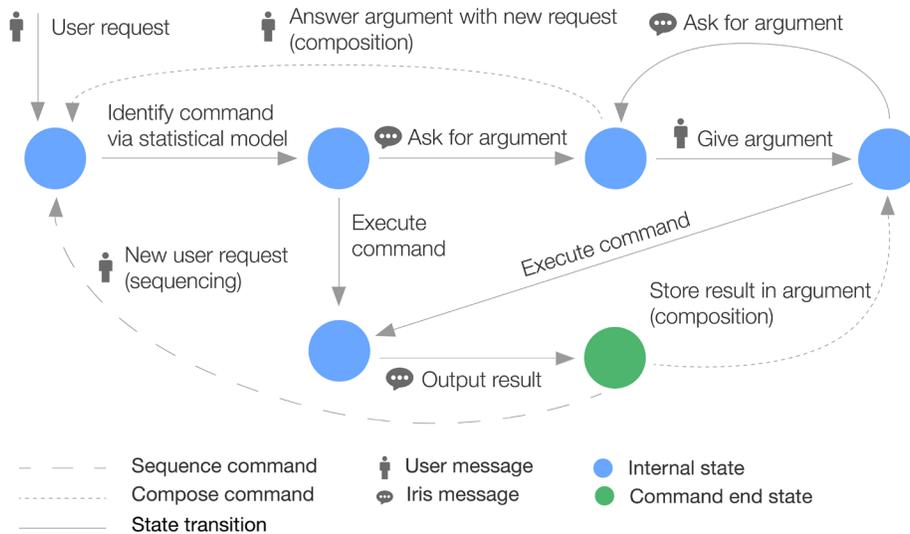

**Figure 3:** Iris allows users to combine commands by transforming them into automata that it can compose and sequence. Here we depict how these automata fit into the system architecture. *Composition* (short dash) allows comands to be called recursively within each other. *Sequencing* (long dash) allows commands to happen in series, referencing previous command results.

This automaton transitions to another that *gathers missing arguments*, cycling through clarification requests with the user (driven by argument type). This in turn transitions to an *execution* automaton that looks up the argument bindings and runs the code.

**The Conversational Type System**

Iris's conversational type system is critical for managing user interactions. First, it provides a framework for resolving a command's arguments through conversation. For example, if Iris knows it needs an Int, it can execute conversational logic to collect that value, such as, "What integer would you like to use for n?". Second, the type system provides guardrails (i.e., dynamic type-checking) that prevent Iris from failing when users ask it to do something that would not work, or when it misinterprets a user request. Third, the type system provides guidance that allows Iris to do the right thing when a user provides a response that—if interpreted literally—would not make sense, but can be transformed to do so. For example, providing a data collection when asked for an array, or documents when asked for a matrix of features.

Types in Iris consist of (1) logic that defines what Python values match the type, (2) a means of converting user input into a value of the type, such as a function to convert the string value "9" into an integer, or a function that looks up the value of a variable name, (3) a clarification request that determines how Iris will interact with a user to resolve a type, such as "what integer would you like for n?", and (4) a set of type converters that can transform values of non-matching types into the correct type, such as converting a collection of data into an array by selecting a column from those data.

Iris currently contains types for *integers*, *strings*, *arrays*, *collections* (multi-dimensional data with named columns), *models* and *metrics* (based on the sklearn API), and *plots* (based on the matplotlib API). These types all inherit from a base class (`IrisType`) that transforms types into automata via the DSL, so an expert user can add new types to the system without worrying about state transition logic. For example, the type definition for integers is only 8 lines of code.

**A Statistical Model of User Requests**

Iris connects user language with commands through a multi-class logistic regression model that is trained to predict commands from user language. Initially, the model is trained based on the examples associated with each Iris command. As a user enters new requests for Iris, the model is updated to incorporate them. As Iris gains more data, it is possible to swap out this statistical model with something more sophisticated, such as a deep neural network [40]. In this paper, the classifier Iris uses is important primarily to enable the other, novel parts of its architecture.

**Executing Requests via Iris Commands**

To connect user requests with commands, Iris acts much like an interpreter in a traditional programming environment. Upon receiving a request, Iris calls its statistical model to determine which command the user wants. Iris then executes

```python
class PearsonCorrelation(IrisCommand):
    title = "compute pearson correlation: {x} and {y}"
    examples = [ "pearson correlation between {x} and {y}",
                 "pearson correlation {x} {y}",
                 "how are {x} and {y} correlated" ]
    help_text = [ "..." ]
    argument_types = {
        "x": Array("Where is the first array to analyze?"),
        "y": Array("Where is the second array?"),
    }
    def command(self, x, y):
        from scipy.stats import pearsonr
        return pearsonr(x,y)
    def explanation(self, corr_pval):
        corr = round(corr_pval[0],4)
        pval = round(corr_pval[1],4)
        return "Correlation of {} with p-value of {}".format(corr, pval)

pearsonCorrelation = PearsonCorrelation()
```

Figure 4: Implementation of an Iris command from a *scipy* function. Iris extends the function with conversational affordances, such as argument requests (lines 8-9) and an explanation (lines 14-17). These affordances translate into automata powered by the conversational DSL.

the desired command's automata, which handle argument extraction and clarification requests.

*Extracting Arguments from Requests*

The automata for argument extraction use an exact match procedure that aligns requests word-by-word with a set of templates and extracts values from matching words in the template so long as they can be converted into the correct type. As users enter requests it has not seen, *Iris learns new templates for executing commands and updates its statistical model*. When arguments cannot be extracted, these automata transition to others that handle clarification requests.

*Composing Command Execution*

Composition allows users to nest commands within each other through conversation (Figure 2A). For example, when Iris asks "what array would you like to analyze?", a user can respond with a request that requires another command to execute, such as "the column with the largest variance".

To achieve composition, automata that handle clarification requests must distinguish between responses that are intended as primitive values (e.g., an integer, or the name of an array value) and those that correspond to new commands. These automata first use the conversion methods provided by their types to attempt to parse a user response into a value (e.g., an array). If this process fails, the automata will process the input as a new, composed command. Iris's interpretation of an input is transparent to the user and updated in real-time as a hint above the text field.

To compose one command's automata (the child) with another's (the parent), the parent initializes the child with a new scope to prevent argument bindings from overwriting each other —for example, when a command is called within itself—passes control of the conversation to the child, and adds a state transition from the child to the itself. After the child executes and transitions to the parent, the parent binds the child's result to the original argument in question, then continues its own execution (Figure 3).

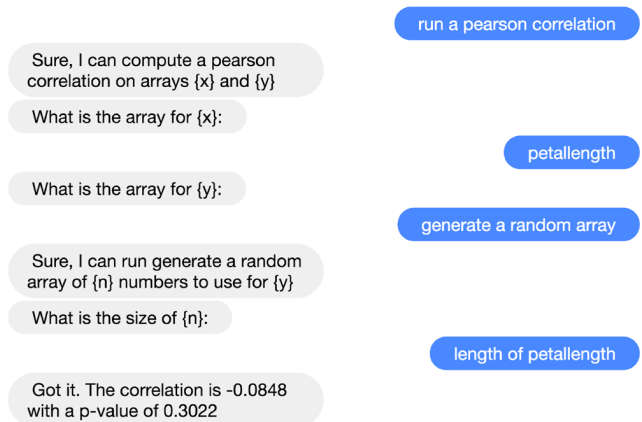
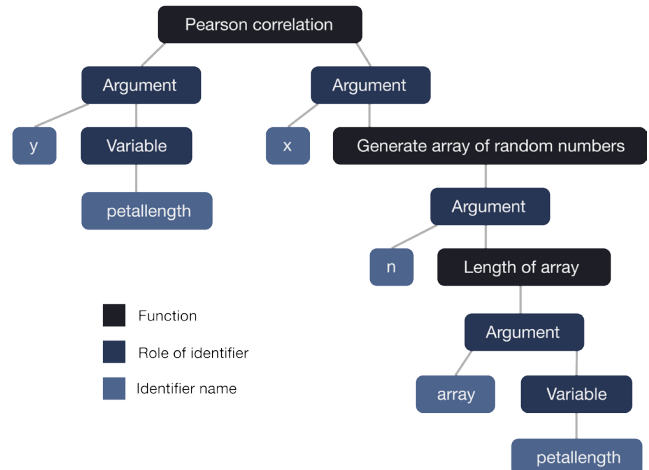

Figure 5: Conversations with Iris are programs under composition: (A) user conversation with Iris, (B) the abstract syntax tree that Iris builds at run-time, and (C) Python code generated by Iris from the AST.

*Sequencing Command Execution*
Sequencing allows users to reference the value of a previous command in their current interaction. To accomplish this, Iris saves the result of each command as it executes in a history variable that can be referred to in future conversations through pronoun keywords such as "this", "those", or "that". For example, after a command has just returned an array, users can ask Iris to "take the mean of that". This simple model works well in practice; in the future, it is possible to swap in more advanced co-reference models [33].

Iris also provides an internal API that commands can call to add new, named variables to the Iris runtime environment. We have used this API to create an general purpose command to `save {value} to {variable name}`, which can be issued to save conversation results for future use, such as in the request, "save that to array result", and then a follow-up, "tell me the variance of array result".

**Transforming Conversations into Programs**
Users can export conversations with Iris as a Python script to save and replicate their work. To accomplish this, Iris incrementally builds an abstract syntax tree (AST) that represents the current state of the conversation. When a user asks to export their conversation, Iris uses rules associated with each AST node type to compile the AST into a Python program. Concretely, Python source code for each relevant command is generated at the top of the program, sequenced commands are translated as variable assignment, and nested commands are translated as function composition (Figure 5C).

**The Iris User Interface**
Users interact with Iris through a chat window that is augmented with hints and metadata about the current state of the system (Figure 1). Users enter requests into an input box (1.1) and receive real-time feedback about what command their re- quest will execute (1.2), which they can use to reformulate the request if necessary. Once a user has entered a command, Iris will begin a conversation in the chat window, asking questions if it needs information about a command's arguments (1.3). Users can respond to these questions with concrete values (for example, "Elena" or "2"), new commands that Iris should execute (1.4) or references to the results of previous conversations (1.5). The chat window has a right sidebar that provides information about the names, types, and values of existing environment variables.

Because Iris is designed for data science tasks, it must also display non-textual data. In particular, Iris outputs complex array and matrix data using *numpy*'s string formatting tools, pretty prints Python objects such as lists and dictionaries, and can embed images directly in the chat window.

The Iris user interface is built in JavaScript with React and Redux and communicates with a Python backend that runs the automata-based logic encoded by the conversational DSL. All backend and frontend components of the user interface are open source at **http://anonymized**.

**EVALUATION**
Can Iris help users accomplish data science tasks? What benefits and drawbacks does a conversational interface provide over programming? We ran a study to validate Iris's design, and compare it to the high-level data mining API provided by the Python package sklearn. Following the study, we asked participants about their impressions of the system.

**Method**
Our study aims to compare Iris to how a data scientist would accomplish a predictive modeling task in Python. Study participants built and cross-validated a model to predict a flower's species based on measurements of the length and width of its petals and sepals [2], then reported on feature relationships by examining model coefficients. This description assumed familiarity with standard modeling terms and procedures (e.g., "logistic regression" and "cross-validate"). The task was written by a third party expert who was not familiar with the training examples for Iris commands. We measured the time it took participants to complete the task, as well as the correctness of their final output.

*Experimental design*: Participants completed the task using both sklearn (in a Jupyter notebook) and the Iris conversational interface, in random order to counterbalance learning effects. We chose to compare to sklearn due to the one-to-one correspondence between the commands in each tool (many Iris commands are built on sklearn).

*Participants*: 8 people participated in the study. All were trained computer scientists with past experience in data science and sklearn, and none had used Iris prior to the study.

*Data format*: In the sklearn condition, we presented the flower data as a dictionary (with column names such as "sepal-length" as the keys). In the Iris condition, we presented these data as a collection (indexed on the same column names). Participants did not need to manipulate the data in either condition, besides selecting features to input to a model.

*User instruction*: In the sklearn condition, we provided a high-level explanation of how the library works, and URLs for all the sklearn methods that a participant would need to complete the task. We also answered any questions that came up in the course of the task (e.g., "what is the cv parameter of cross_val_score?"). In the Iris condition, we explained how the interface worked, but did not provide details about the specific commands that the participant would need, or answer questions about the purpose of Iris commands.

### Results

Participants completed the task 2.57 times faster on average in the Iris condition (Figure 6). This effect was statistically significant under a Mann-Whitney U test ($U_8=3.0$, $p<0.01$). There was a small learning effect (participants were 1.31 times faster on average for the second interface), re-emphasizing the importance of our randomization, but the effect was not significant ($U_8=26$, $p=0.28$). All participants completed the task correctly in both conditions.

Participants approached the modeling task in different ways. Some participants worked backwards from a high level goal that was several steps in the future. For example, P1 asked Iris to "cross-validate" before they had created a model or selected features. When Iris then asked, "what model do you want to use?", P1 created it on demand though a nested conversation (made possible by *composition*). Other participants worked more incrementally. For example, P4 first selected features from the flower data and saved them in named variables, then referenced those variable names in conversation when creating a logistic regression model (made possible by *sequencing*). Most participants took a middle ground approach that leveraged both kinds of combination: for example, composing model creation and data selection, then sequencing that with a new request for cross-validation.

Participants also differed in the vocabulary they used to interact with Iris. For example, P1 asked Iris for a "logistic regression", while P3 asked to "build a classifier", and P2 asked to "select columns" while P5 asked to "select fea-tures". Sometimes users reformulated a request when nothing appropriate appeared in the hint box, as occurred when one participant attempted to access a column of data using the '.' notation common to Python objects.

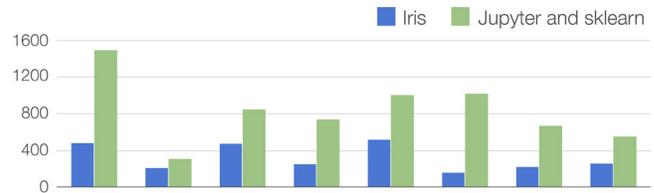

Figure 6: Study participants completed a data science task 2.57 times faster on average with Iris than with sklearn and Jupyter ($p < 0.01$). The y-axis describes time in seconds.

While three participants interacted with Iris almost entirely through full natural language queries, most interacted via keywords, as if they were querying a search engine (for example, typing "logistic regression" to trigger `create a classification model`). When asked why they chose to converse this way, participants said it was faster, as the hint box indicated when keywords would trigger the correct command. Notably, keyword searches still allowed participants to build complex commands through composition and sequencing. In contrast, participants who interacted with Iris through full language queries (for example, "cross-validate model1 with accuracy and 10 folds") reported that they wanted Iris to extract a command's arguments automatically.

### Advantages to Conversation

In interviews following the study, participants mentioned aspects of Iris that they liked and disliked. These aspects varied to some degree with a user's level of experience with sklearn. For example, less experienced users of sklearn found value in the *structural guidance* provided by conversation with Iris (for example, how APIs fit together). P1 said:

> "It took a while to remember how to thread together the components of the sklearn API. Like, first you need to initiate a model class instance, then pass that to the cross-validation function—it wasn't totally obvious. But with Iris, once it understood I wanted to cross-validate, it walked me backwards though the steps, like giving it a model, what type of model, and so on."

In contrast, more experienced users thought Iris would *save time* as a wrapper for a set of functions they frequently call in smaller scripts. For example, P4 said:

> "For simple scripts, this is so much faster... I really like that you don't have to remember the name of the function you want to call. All you need to do is say something close, like 'rename column' and it can figure out the rest."

Other participants commented on the *complexity* of what the system could accomplish. For example, P3 said:

> "It's so cool that Iris can do such complex stuff, when legit companies out there have nothing this sophisticated."

Four participants expressed interest in using Iris for future data science work. One participant offered to connect us with a colleague in the Psychology department who was teaching an introductory statistics class as applied through R. "Iris would be so much better," P4 said.

**Challenges for Conversation**

We also asked participants what they found challenging about Iris's conversational interface. One common view was that the expressivity of Iris presented more *opportunities for mistakes*. For example, P2 said:

> "It's great that Iris can combine anything together, but that creates more opportunities for something to go wrong."

Iris's conversational type system can prevent most common user errors, such as combining commands that return incompatible types, or answering Iris with the wrong type of data, but it is not a panacea. One of our goals when designing Iris was to ensure that—whatever input a user enters—the system will be robust to both system and user mistakes, as strong guardrails can encourage user exploration and experimentation [30].

Similarly, other kinds of mistakes can emerge from miscommunication between a system and user about what the system is actually doing. Along these lines, some participants expressed the concern that a natural language based system should be *transparent about the operations it is executing*, especially for data science work. P5 said:

> "Say I have some data that's not normally distributed, and I ask Iris to do a t-test. How do I know it's doing the right one?"

In addition to the fact that Iris repeats back what commands it is executing (for example, "Sure, I can do a Mann-Whitney U test"), which provides user with an immediate sanity check, the system allows users to inspect a conversation's underlying Python code. This code is largely written though high level APIs such as scikit, so it is possible to quickly inspect the constituent functions and determine what methods (such as what kinds of t-tests) are being executed.

Other participants mentioned the *vocabulary problem* as a potential challenge. For example, P4 said:

> "Everything worked for me, but I can imagine another user entering the wrong words and getting stuck."

The vocabulary problem has long been an issue for systems based on natural language. As Iris gains more users and language data, we aim to address this problem more formally by learning from logs of mistakes. In the course of the pilot study, three participants entered requests that triggered the wrong command. For example, "make model" triggered "create a regression model" instead of "create a classification model"; and for two participants, entering the name of a column did not resolve to a command to extract that column from a collection. While these mistakes were foreshadowed by text in the system's hint box, users ignored or misinterpreted these hints, and future versions of the Iris might emphasize them further.

Finally, several participants suggested that *text is not always the best medium* for communicating. For example, P2 said:

> "There are certain things, like selecting data columns, that I'd prefer to do by clicking on things."

Prior work has demonstrated strong results with mixed modality interfaces [21], and we plan to incorporate other modalities into Iris moving forward. Manipulating and transforming data, for example, is perhaps best accomplished through a spreadsheet-like interface, and we plan to explore how such a module might augment user conversation.

**LIMITATIONS AND FUTURE WORK**

Here we discuss challenges to overcome as Iris's user base and range of functionalities grow.

First, as Iris expands to support many more commands, interpreting user language may become more difficult. The system currently supports 95 commands with a manually authored dataset of examples (roughly 5 examples per command). How accurate will command classification become as the set of commands grows larger? How many user examples are necessary to support a new command among a library of thousands of others? Addressing these concerns will become more important as we deploy Iris in the wild.

Second, Iris creates AST representations of conversations with users and so has the ability to save these recorded programs, which may combine multiple commands. This ability connects with work in programming by demonstration (PBD) [25], and offers the potential for users to create complex, reusable workflows through natural language. Saving ASTs for reuse presents a usability challenge, however. For example, how should Iris ask a user which parameters in the AST are arguments, and which should be captured as constants? And is it be possible to mine useful higher-level commands by analyzing the ASTs of hundreds or thousands of users? We aim to explore these questions as Iris grows.

Third, Iris composes commands through turn-taking in conversation. This contrasts with algorithms such as semantic parsing [3], which can learn composition within individual utterances: for example, translating "how old is Elena Ferrante" into `get_age(lookup_person("Elena Ferrante"))`. Unlike Iris, however, these models require large amounts of training data. User conversations with Iris provide one means of collecting such data, and we plan to explore the application of semantic parsing (and other modern NLP) in future work.

Finally, Iris enables exploratory and interactive data analyses, as you might conduct today in R or a Jupyter notebook. This is distinct from other data science work that requires enormous datasets, where training a model may take hours, days, or even weeks. To run these models, researchers typically setup and debug long-running pipelines of commands, which is not something Iris is currently designed to do. As Iris expands to allow users to create and save workflows though PBD, such pipelines may be more feasible to run.

**CONCLUSION**

In this paper, we show how conversational agents can draw on human conversational strategies to combine commands together, allowing them to assist us with tasks they have not been explicitly programmed to support. We showcase these ideas in Iris, an agent designed to help users with data science and machine learning tasks. More broadly, our work demonstrates how simple models of conversation can lead to surprisingly complex emergent outcomes.